\documentclass[nofootinbib,aps]{revtex4}
\usepackage{amssymb,amsmath,graphicx,color,microtype}
\usepackage{hyperref}
\hypersetup{
    colorlinks=true,
    linkcolor=blue,
    filecolor=blue,      
    urlcolor=blue,
    citecolor=blue
}
\usepackage[paperwidth=210mm,paperheight=297mm,centering,hmargin=2cm,vmargin=2.5cm]{geometry}

\begin{document}
\title{On the Effective Field Theory for Quasi-Single Field Inflation}
\author{Xi Tong$^1$}
\email{tx123@mail.ustc.edu.cn}
\author{Yi Wang$^{2,3}$}
\email{phyw@ust.hk}
\author{Siyi Zhou$^{2,3}$}
\email{szhouah@connect.ust.hk}
\affiliation{${}^1$School of Physics, University of Science and Technology of China, \\
Hefei, Anhui 230026, China}
\affiliation{${}^2$Department of Physics, The Hong Kong University of Science and Technology, \\
  Clear Water Bay, Kowloon, Hong Kong, P.R.China}
\affiliation{${}^3$Jockey Club Institute for Advanced Study, The Hong Kong University of Science and Technology, \\
   Clear Water Bay, Kowloon, Hong Kong, P.R.China}

\begin{abstract}
We study the effective field theory (EFT) description of the virtual particle effects in quasi-single field inflation, which unifies the previous results on large mass and large mixing cases. By using a horizon crossing approximation and matching with known limits, approximate expressions for the power spectrum and the spectral index are obtained. The error of the approximate solution is within $10\%$ in dominate parts of the parameter space, which corresponds to less-than-$0.1\%$ error in the $n_s$-$r$ diagram. The quasi-single field corrections on the $n_s$-$r$ diagram are plotted for a few inflation models. Especially, the quasi-single field correction drives $m^2\phi^2$ inflation to the best fit region on the $n_s$-$r$ diagram, with an amount of equilateral non-Gaussianity which can be tested in future experiments.
\end{abstract}

\maketitle

\section{Introduction}

Any realistic inflation model accommodates plenty of massive fields. Those massive fields bring rich new physics \cite{Linde:1993cn,Yamaguchi:2005qm,Chen:2009we,Chen:2009zp,Tolley:2009fg,Achucarro:2010jv,Jackson:2010cw,Cremonini:2010ua,Baumann:2011su,Baumann:2011nk,Achucarro:2012sm,Assassi:2012zq,Sefusatti:2012ye,Norena:2012yi,Chen:2012ge,Pi:2012gf,Achucarro:2012yr,Gwyn:2012mw,Noumi:2012vr,Cespedes:2013rda,Gong:2013sma,Emami:2013lma,Castillo:2013sfa,Kehagias:2015jha,Liu:2015tza,Achucarro:2015rfa,Arkani-Hamed:2015bza,Dimastrogiovanni:2015pla,Maldacena:2015bha,Schmidt:2015xka,Chen:2015lza,Bonga:2015urq,Chen:2016cbe,Chen:2016nrs,Flauger:2016idt,Lee:2016vti,Chen:2016qce,Liu:2016aaf,Delacretaz:2016nhw,Meerburg:2016zdz,Chen:2016uwp,Chen:2016hrz,Jiang:2017nou,Chen:2017ryl,An:2017hlx}. Quasi-single field inflation \cite{Chen:2009we,Chen:2009zp,Baumann:2011nk} aims to study the characteristic features of those massive field perturbations. 

As examples of massive fields, string oscillatory modes and Kaluza-Klein modes of extra dimensions generically generates massive fields with mass $m>H$. While effects of Standard Model uplifting \cite{Chen:2016nrs,Chen:2016uwp,Chen:2016hrz}, supersymmetry breaking \cite{Baumann:2011nk,Delacretaz:2016nhw} and non-minimal coupling generate massive fields with mass $m\sim H$ (up to additional factors from coupling constants). At the homogeneous and isotropic background level, unless otherwise excited, those massive fields stay at their potential minima. However, at the perturbation level, the impact of those massive fields show up through virtual particles which can be described by an effective field theory (EFT) \cite{Tolley:2009fg,Achucarro:2010jv,Achucarro:2012sm,Gwyn:2012mw}, and real particles due to the particle production process of a time-dependent background \cite{Chen:2009we,Chen:2009zp,Arkani-Hamed:2015bza}.

The real particle contribution in quasi-single field inflation predicts characteristic signatures on non-Gaussianities \cite{Chen:2009we,Chen:2009zp,Arkani-Hamed:2015bza}. In the squeezed limit, the massive fields decay to curvature perturbation while being diluted away, producing a family of quasi-local shapes of non-Gaussianities. Especially, when the massive fields are underdamped by Hubble friction, the massive field effects manifest themselves as interference patterns of the quantum phases of massive particles between their creation and decay. The real particle production effects help test high energy physics and understand the primordial universe in many ways, for example
\begin{itemize}
  \item As quantum primordial standard clocks \cite{Chen:2015lza,Chen:2016cbe,Chen:2016qce} (see also \cite{Chen:2011zf,Chen:2011tu,Chen:2014cwa} for classical primordial standard clocks). Conventionally, from the scales $k$ of the observed density fluctuations, one can infer the physics of Hubble-crossing in the primordial universe, and thus the physics happened at conformal time $\tau = -1/k$. However, without the trace of physical time, the information about the expansion history of the universe is missing. This is why the current high-precision observation may still leave space for the study of alternative-to-inflation scenarios \cite{Khoury:2001wf,Brandenberger:1988aj,Wands:1998yp,Finelli:2001sr}. It is crucial to note that, the mass of a field is a physical parameter, which provides direct information about the physical time of the primordial universe (for example, from the physical oscillation frequency of the massive field). Once the physical time as a function of conformal time is known, its differentiation gives the scale factor $a(t)$ of the primordial universe.  Thus, observing the massive field interference patterns in non-Gaussianities is a direct way to measure the expansion history of the primordial universe. Once observed, such non-Gaussian correlations provide a proof of inflation (or a proof of an alternative-to-inflation scenario). 
  \item As a cosmological collider \cite{Arkani-Hamed:2015bza,Chen:2016nrs,Chen:2016uwp,Chen:2016hrz}. The characteristics of the massive particle, namely, mass, spin and coupling are encoded as different features of non-Gaussianities -- its behavior in its isosceles squeezed limit, non-isosceles squeezed limit and the absolute amplitude, respectively. Thus in principle, all the massive particle spectra is recorded in the density fluctuation correlations of the universe. And the challenge is how many modes we can access and how well we can deduct the late-time sources of non-Gaussianities. Once those massive particles are detected during inflation, they provide invaluable hints on the future development of particle physics.
  \item As a test of quantum mechanics in the primordial universe \cite{Maldacena:2015bha}. During inflation, the light degrees of freedom, namely the inflaton and the graviton, transits from its quantum vacuum to a classical state within a very short time (horizon crossing). Thus they are not ideal for studying the quantum nature in the primordial universe, such as entanglements, Bell inequalities, etc. Massive fields have slowly increasing or non-increasing particle number and thus provide a cleaner arena to study the quantum effects of the primordial universe \cite{Liu:2016aaf}.
\end{itemize}

As encoded in the fundamental principles of quantum mechanics, the above real particle effects always accompany virtual particle effects. It is also important to study the virtual particle effects of quasi-single field inflation in fine details, because
\begin{itemize}
  \item As circumstantial evidences, the virtual particle effects can better confirm the real particle effects, to promote the confidence level of future observations.
  \item Sometimes the real particle effects are too small to be observed. In this case, the virtual particle effects provide systematic errors in determining inflation models from observations. We should understand those systematic errors \cite{Achucarro:2015rfa,Jiang:2017nou}.
\end{itemize}

In this paper, we focus on the virtual particle effects. We study the case where the inflaton fluctuation and the massive field are coupled by
\begin{align}\label{operator}
  \Delta \mathcal {L} \propto \rho \dot\pi \sigma ~,
\end{align}
where $\pi$ is the (canonically normalized) Goldstone of time translation symmetry breaking, reflecting the inflaton fluctuation, $\sigma$ is the massive field, and $\rho$ is a mixing parameter with mass dimension. Theoretically, this mixing term can originate from a Lorentz invariant dimension-5 operator
\begin{align}\label{operator5}
  {\mathcal O}_5 = - \frac{1}{2\Lambda} (\partial\phi)^2 \sigma ~.
\end{align}
This dimension-5 operator reflects the UV sensitivity of inflationary perturbations. This is because $\rho \propto \dot\phi / \Lambda$. And $\dot\phi = H^2 / (2\pi P_\zeta^{1/2}) \simeq 3000 H^2$. In other words, described by an EFT, the inflationary fluctuations can get contributions from this dimension-5 operator if new physics arises at $\Lambda \lesssim 3000 H$. Especially, if $H\sim 10^{13}$GeV (which implies observable primordial gravitational waves in the next generation experiments), then the new physics scale is about $10^{16} \sim 10^{17}$GeV. This is well motivated in grand unification and perturbative string theory.

The inflationary power spectrum from Eq.~\eqref{operator} has been extensively studied in the literature. The weak coupling case $\rho \ll m$ (where $m$ is the mass of $\sigma$) is numerically studied in \cite{Chen:2009zp} and the analytical result is obtained in \cite{Chen:2012ge}. In the large mass limit, an EFT is studied in \cite{Tolley:2009fg,Achucarro:2010jv,Achucarro:2012sm}. The strongly coupled regime is studied in \cite{Baumann:2011su,Cremonini:2010ua,Gwyn:2012mw,An:2017hlx}. In \cite{Gwyn:2012mw}, an improved EFT is proposed, which applies for both cases. But to date, an detailed analysis for the implication of this EFT (for all parameters satisfying $m^2+\rho^2 \gg H^2$) is missing. In this paper, we fill this gap and study this EFT in the whole parameter regime where an EFT description is applicable. As checked numerically, the EFT still works reasonably well even when $m^2+\rho^2 \sim H^2$.

This paper is organized as follows: In Section~\ref{sec:eft}, we discuss the improved EFT approach. In Section~\ref{sec:numerical}, we provide a method to numerically solve the single field EFT. In Section~\ref{sec:numerical}, we discuss the impact of massive fields on power spectra observables, namely the $n_s-r$ diagram. We conclude in Section~\ref{sec:conclusion}.

\section{The Improved Effective Field Theory}
\label{sec:eft}

Considering the two-point mixing operator~\eqref{operator}, we start from the following free action of the quasi-single field inflation written in terms of the conformal time $\tau = - e^{-Ht}/H$,
\begin{align}
S[\pi,\sigma] = \int d^3 x d \tau \frac{1}{2 H^2\tau^2} \bigg[ (\partial_\tau \pi)^2 - (\nabla \pi)^2 +(\partial_\tau \sigma)^2 - (\nabla \sigma)^2 -\frac{m^2}{H^2\tau^2} \sigma^2 -\frac{2\rho}{H\tau} \sigma \partial_\tau \pi \bigg]~.
\end{align}

If we do not treat the two-point coupling between the two fields as interaction, then by the standard way to canonically quantize the system, we can write the fields in the conformal momentum space as 
\begin{align}
\pi_{\mathbf k} (\tau) = u^{(1)}_{k} (\tau) a^{(1)}_{\mathbf k} + u^{(2)}_{k} (\tau) a^{(2)}_{\mathbf k}  + {\rm h.c.} ~,\\
\sigma_{\mathbf k} (\tau) = v^{(1)}_{k} (\tau) a^{(1)}_{\mathbf k} + v^{(2)}_{k} (\tau) a^{(2)}_{\mathbf k}  + {\rm h.c.}~.
\end{align}
where the superscript (1) and (2) denote two different sets of modes defined in \cite{An:2017hlx}. This leads to a set of coupled differential equations
\begin{align}\label{couple}
 {u''_{k}}  -  \frac{2 u'_k}{ \tau} + k^2 u_k - \frac{\rho}{H} \left( \frac{v'_k}{ \tau} - \frac{3 v_k}{  \tau^2} \right) =0 ~,\\
 {v''_k} - \frac{2 v'_k }{ \tau} + \left( k^2 + \frac{m^2 }{H^2   \tau^2} \right) v_k + \frac{\rho}{H} \frac{u'_k}{ \tau} =0 ~.
\end{align}
the prime $'$ denotes $\partial_\tau$, the derivative with respect to the conformal time. It is possible to solve these coupled differential equations numerically. The initial conditions are determined by looking at the correctly normalized solutions to equations of motion in the highly UV regime \cite{An:2017hlx}. This coupled differential equation is in general difficult to solve analytically. However, if we are only interested in the virtual particle contribution, we can integrate out the $\sigma$ field. Due to the special form of the interaction, we can view the whole Lagrangian as a quadratic polynomial and a Gaussian integration can be directly performed. In the following, we study an improved EFT, which can take into account both the local and non-local contributions of the virtual particle effects. After we integrate out the heavy field, we are left with only the Goldstone field. It is widely known that during inflation, this field will freeze at horizon crossing. We can further take advantage of this nice property to have an approximate method to solve for the power spectrum.  

In general, the path integral formulation of a time dependent quantum system is realized by the Schwinger-Keldysh formalism \cite{Chen:2017ryl}, where two sets of field variables need to be taken into account, one time-ordered and the other anti-time-ordered. Here because we focus only on the form of the action, we only consider one of the two (for example, the time-ordered one). Thus the generating functional with zero external field is,
\begin{align}\nonumber
Z & = \int \mathcal D \pi \mathcal D \sigma \exp \bigg( {i \int d^3 x d \tau} \mathcal L \bigg) \\ \nonumber
& =  \int \mathcal D \pi \exp \bigg( {i \int \frac{d^3 x d\tau}{2 H^2 \tau^2} ((\partial_\tau\pi)^2-(\nabla \pi)^2) } \bigg) \int \mathcal D \sigma \exp \bigg(i \int \frac{d^3 x d\tau}{2 H^2 \tau^2} [(\partial_\tau \sigma)^2 - (\nabla \sigma)^2 - \frac{m^2 }{H^2\tau^2} \sigma^2 - \frac{2\rho}{H\tau} \sigma\partial_\tau \pi  ] \bigg) \\
& = \int \mathcal D \pi \exp \bigg( i S_{\rm eff} [\pi] \bigg)~.
\end{align}  
When performing the path integral, we change variable to the canonically normalized field since this matches the flat space field in the UV limit. We define a canonically normalized field $\varsigma \equiv a \sigma$. Assuming the integration measure is unchanged, the $\sigma$ part of the integral can be written in terms of $\varsigma$ as
\begin{align}
 \int \mathcal D\mathcal \varsigma \exp \bigg( i \int \frac{d^3 x d\tau}{2} [\varsigma \Box \varsigma + 2 \rho a^2 \varsigma \partial_{\tau} \pi + (\rho a^2 \partial_\tau \pi) \frac{1}{\Box} (\rho a^2 \partial_\tau \pi) - (\rho a^2 \partial_\tau \pi)\frac{1}{\Box} (\rho a^2 \partial_\tau \pi)  ] \bigg)~.
\end{align}
where
\begin{align}
 \Box \equiv - \partial^2_\tau + \nabla^2 - (m^2 a^2 - 2 a^2 H^2 )~.
\end{align}
Integrating out $\varsigma$, we obtained the effective action for the $\pi$ field,
\begin{align}
Z =  \int \mathcal D \pi \exp \bigg\{ i \int \frac{d^3 x d\tau}{2 H^2 \tau^2} \bigg[  (\partial_\tau\pi)^2-(\nabla \pi)^2   +  a^{-2} \rho^2 \pi \partial_\tau \bigg(a^2 \frac{1}{\Box} a^2 \partial_\tau \bigg) \pi  \bigg] \bigg\} ~.
\end{align}
After Fourier transform $\nabla \rightarrow i k$, so $(\nabla\pi)^2  \rightarrow (i k \pi ) (-i k \pi) = k^2 \pi^2$, we have the  operator in momentum space 
\begin{align}\label{box}
\Box = - \partial^2_\tau - k^2 - (m^2 a^2 - 2 a^2 H^2 ) ~.
\end{align}
The real particle production is related to the situation where $\Box = 0$, which means there exists a pole in the propagator. Another way of explanation is that the intermediate $\sigma$ particle goes on-shell. Since in the effective picture we do not see real $\sigma$ particles, this is the place where the effective theory breaks down and particle production of $\sigma$ happens. As we check below, the real particle production is indeed subdominant and thus can be neglected concerning the calculation of power spectrum. The second term in ~\eqref{box} characterizes the non-local contributions. In the large $\rho$ limit, this term has the dominant contribution. The third term characterizes the local contributions. In the large $m$ limit, this term dominates.
Since we try to incorporate all the effects brought about by $m$ and $\rho$ when there's no $\sigma$ particle produced, we want to consider the following expansion for the propagator
\begin{align}
\frac{1}{\Box}   = - \frac{1}{k^2 + (m^2 a^2 - 2 a^2 H^2)} + \frac{\partial^2_\tau }{ (k^2 + (m^2 a^2 -2 a^2 H^2))^2 } +\ldots~.
\end{align}
The first term characterizes all the contributions from the virtual particles, the second order term and higher order terms contribute to real particle production.
Inserting only the first term into the effective action, we obtain
\begin{align}\label{eff}
S_{\rm eff} = \frac{1}{2} \int \dfrac{d^3 k}{(2\pi)^3} d\tau \frac{1}{H^2\tau^2} \bigg[ \left( 1 + \frac{\rho^2}{k^2 H^2 \tau^2 + m^2 - 2 H^2} \right) \pi'^2 - k^2 \pi^2 \bigg]~.
\end{align}
This action is similar to that obtained in \cite{Gwyn:2012mw}. There exist comprehensive studies of this effective field action in the large $\rho$ limit \cite{Baumann:2011su,Cremonini:2010ua,Gwyn:2012mw,An:2017hlx} and large $m$ limit \cite{Tolley:2009fg,Achucarro:2010jv,Achucarro:2012sm}, here we briefly review the results in the Appendix~\ref{largeEFT}.

With the presence of a sound speed, the time a mode crosses the horizon changes as $-k\tau \sim c_s^{-1}$ because this time the perturbations travel in the conformal Minkowski spacetime at a subluminal sound speed $c_s<1$. After sound horizon crossing, the mode will freeze and have no super-Hubble propagation. The massive field has a characteristic time scale $-k\tau \sim m/H$. When this characteristic time scale appear after horizon crossing, it cannot change the power spectrum. Thus the large $\rho$ EFT is applicable when $c_s^{-1}\sim \sqrt {\rho/H}\gg m/H$. In a similar way, the large $m$ effective field theory is applicable when $\sqrt{\rho/H} \ll m/H$. In the following, we present our result for the improved EFT which can accommodate these two results and can deal with the situation that the characteristic time scale of the massive field coincides with the sound horizon crossing scale.
 
From the effective action~\eqref{eff}, we can read off an effective sound speed as
\begin{align}\label{dispersion}
c_s^{-2} (k\tau) = 1 +  \frac{\rho^2}{k^2 H^2 \tau^2 + m^2- 2 H^2} ~.
\end{align}
This is previously derived as a dispersion relation in \cite{Gwyn:2012mw}. One can estimate $\tau$ using the horizon crossing formula
\begin{align}
k\tau = B  c_s^{-1}~,
\end{align}
where $B$ is a constant to be determined by matching with the large $\rho$ EFT. After horizon crossing, the mode freezes. We insert this expression into ~\eqref{dispersion} and obtain the equation
\begin{align}
c_s^{-2}   = 1 +  \frac{\rho^2}{H^2 B^2 c_s^{-2} + m^2 - 2 H^2} ~.
\end{align}
There are four solutions, we take the real positive solution for the inversed sound speed
\begin{align}
c_s^{-1} = \sqrt{\frac{2(\rho^2+m^2- 2 H^2)}{\sqrt{4B^2H^2\rho^2+(B^2H^2+m^2- 2 H^2)^2}-B^2H^2+m^2- 2 H^2}}~.
\end{align}
In the large mass limit $m^2\gg \rho H$, this expression becomes
\begin{align}
c_s^{-1} = \sqrt{1+\frac{\rho^2}{m^2- 2 H^2}}~.
\end{align}
In the large $\rho$ limit $m^2\ll \rho H$, this expression becomes
\begin{align}
c_s^{-1} = \sqrt{\frac{\rho}{BH}} ~.
\end{align}
Matching this one with the large $\rho$ expression
\begin{align}
\sqrt{\frac{\rho}{BH}} = \mathcal C \sqrt{\frac{\rho}{H}}~.
\end{align}
We obtain $\sqrt B = \mathcal C^{-1}$, $\mathcal C =\dfrac{16\pi}{\Gamma(-1/4)^2}\sim 2.09$ is obtained in \cite{Gwyn:2012mw,An:2017hlx}. Note that we have assumed here that the factor $B$ does not depend on $\rho$. This is the main source of error (about $10\%$) in this analytical method. The power spectrum is now corrected as
\begin{align}
P_{\zeta} = P_{\zeta}^{(0)} c_s^{-1},\quad 
c_s^{-1} =  \sqrt{\frac{2 \left(m^2 - 2 H^2 +\rho ^2\right)}{ m^2 - 2 H^2-\frac{H^2}{\mathcal{C}^4}+\sqrt{\left( m^2 -2 H^2 +\frac{H^2}{\mathcal{C}^4}\right)^2+\frac{4
   H^2 \rho ^2}{\mathcal{C}^4}}}}  ~.
\end{align} 
This result is obtained by physically considering sound horizon crossing can recover the result of large $\rho$ EFT in the limit $\rho \rightarrow \infty$ and the result of large $m$ EFT when $m\rightarrow \infty$. Moreover, no hierarchy between the two scales $\sqrt{\rho/H}$ and $m/H$ is needed.

\section{Numerical Solution of the EFT Equations}
\label{sec:numerical}
In order to test the improved EFT method as well as the assumption that an EFT can apply at all, we should compare our result with the numerical result in the full theory. Using the variation principle of the effective action~\eqref{eff}, we can obtain the equation of motion of the $\pi$ field,
\begin{align}\label{eom}
\pi '' - \frac{2}{\tau} \pi' \left( 1+ \frac{\rho^2 H^2 k^2 \tau^2}{(m^2- 2 H^2 + H^2 k^2 \tau^2)(m^2- 2 H^2 + H^2 k^2 \tau^2 + \rho^2)} \right) +  k^2  \frac{1}{1+\frac{\rho^2}{k^2 H^2 \tau^2 + m^2- 2 H^2}} \pi = 0~.
\end{align} 
We can see that after deriving the equation of motion, we got an correction to the Hubble parameter, which is in the bracket of the second term. The third term corresponds to the change of the sound speed which we previously obtained. Numerically solving this equation is proved to be challenging and it turns out difficult to obtain an accurate result. Here we avoid this problem by casting this equation into an equation in the form of a harmonic oscillator with time-dependent frequency. Make the following variable substitution
\begin{align}
\chi = \frac{\sqrt{m^2- 2 H^2+ \rho^2 + H^2 k^2 \tau^2}}{\tau \sqrt{m^2- 2 H^2 +  H^2 k^2 \tau^2}} \pi~.
\end{align} 
With the new $\chi$ field, we can eliminate the first order derivative term which previously exists in the equation of motion of the $\pi$ field. The equation of motion for the $\chi$ field is
\begin{align} \label{eq:numeq}
\chi'' + w^2(\tau) \chi = 0~,
\end{align}
where the coefficient $w$ is
\begin{align}\nonumber
w^2(\tau) =k^2 \left(c_s^2-\frac{H^2 \rho ^2 \left(5 H^4 k^4 \tau ^4+2 H^2 k^2 \tau ^2 \left(-6 H^2+3 m^2+2 \rho ^2\right)+\left(m^2-2 H^2\right) \left(-2 H^2+m^2+\rho
   ^2\right)\right)}{\left(H^2 \left(k^2 \tau ^2-2\right)+m^2\right)^2 \left(H^2 \left(k^2 \tau ^2-2\right)+m^2+\rho ^2\right)^2}\right)-\frac{2}{\tau ^2}~.
\end{align}
Eq.~\eqref{eq:numeq} is then solved numerically. The results of the numerical calculation are be represented in the below section in terms of power spectra observables.

Notice that this method is only applicable in the parameter region $m>\sqrt{2}H$ because there exist a singularity in Eq.~\eqref{eq:numeq} if $m<\sqrt{2}H$. So in this region we still use the original Eq.~\eqref{eom}, where instead of a singularity, we only have a stiff point in the equation of motion. 

\section{Impacts on Power Spectra Observables}
In this section, we compare the improved EFT (both analytical approximation and numerical solution) with the numerical solution of the full two-field theory. to see how well our result can approach the real power spectra observables.  

\begin{figure}[htbp] 
  \centering 
  \includegraphics[width=0.8\textwidth]{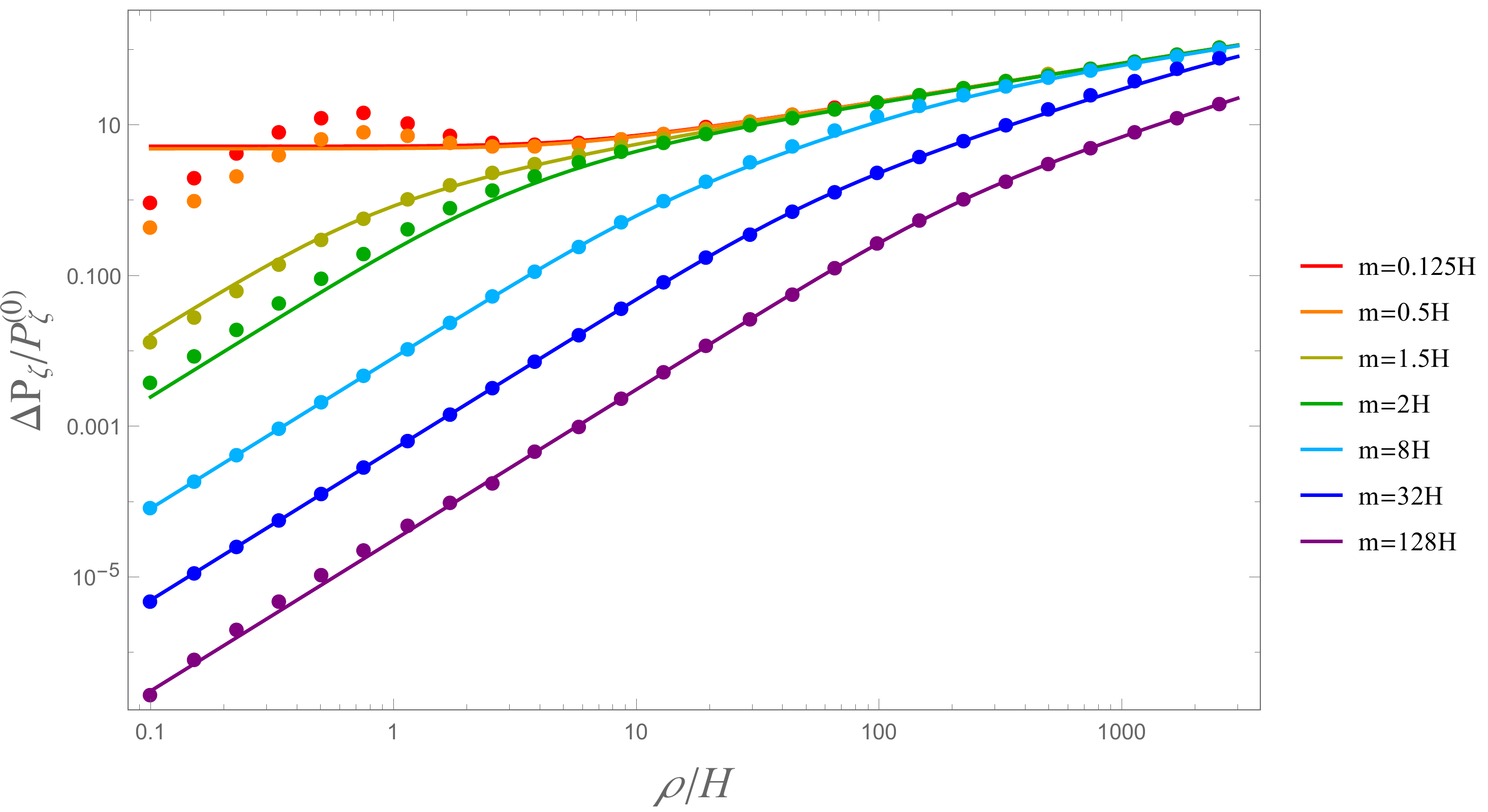}   
  \caption{ The correction to the primordial power spectrum. The dots are results from numerically solving the coupled differential equations of the $\pi$ field and $\sigma$ field. The solid line is our result. In the parameter region $m^2+\rho^2>4H^2$, our new analytical result agrees well with the numerical result. The red, orange, yellow, green, cyan, blue and purple corresponds to $m=0.125H,0.5H,1.5H,2H,8H,32H,128H$, respectively.} 
 \label{fig:Pzeta}
\end{figure}

We first look at the correction to the primordial power spectrum. We compare the improved EFT results and that of numerically solving the coupled differential equations in the full theory (See Fig.~\ref{fig:Pzeta}). The improved EFT agrees well with numerically solving the coupled differential equations when the mass $m$ and the coupling $\rho$ satisfy $m^2 + \rho^2 > 4 H^2$.

\begin{figure}[htbp]
    \centering
    \begin{minipage}{0.7\textwidth}
        \centering
        \includegraphics[width=0.8\linewidth]{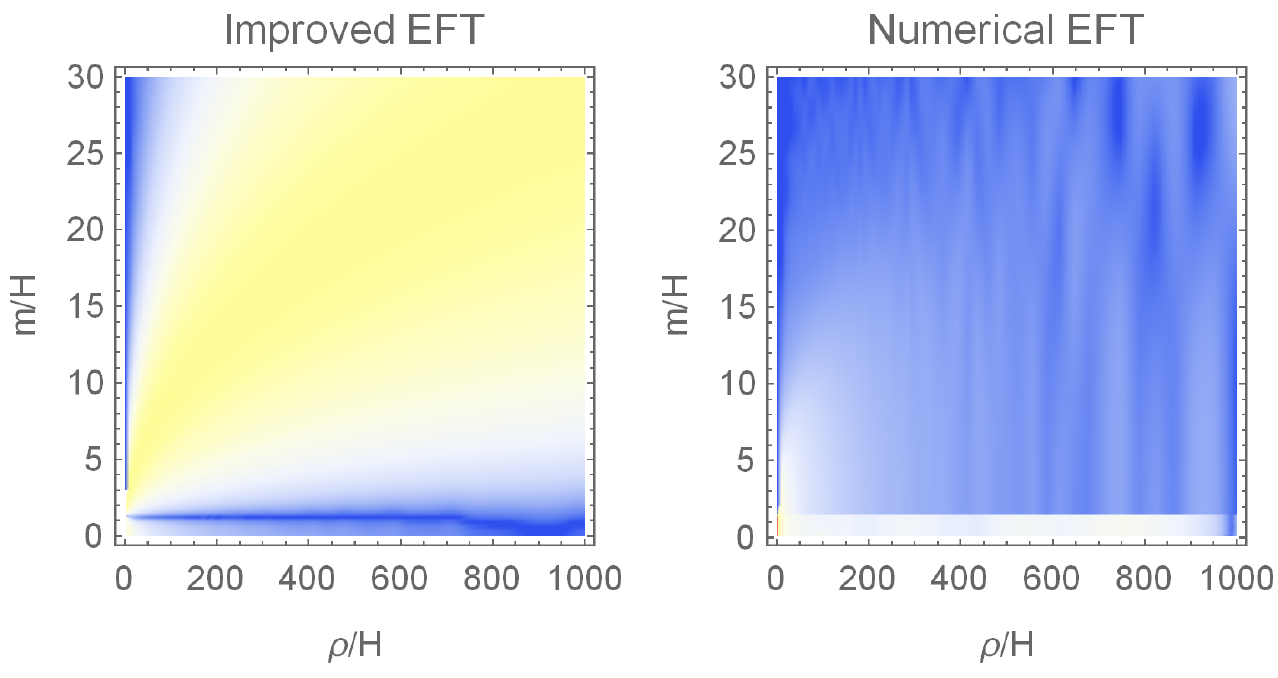} 
        \includegraphics[width=0.8\linewidth]{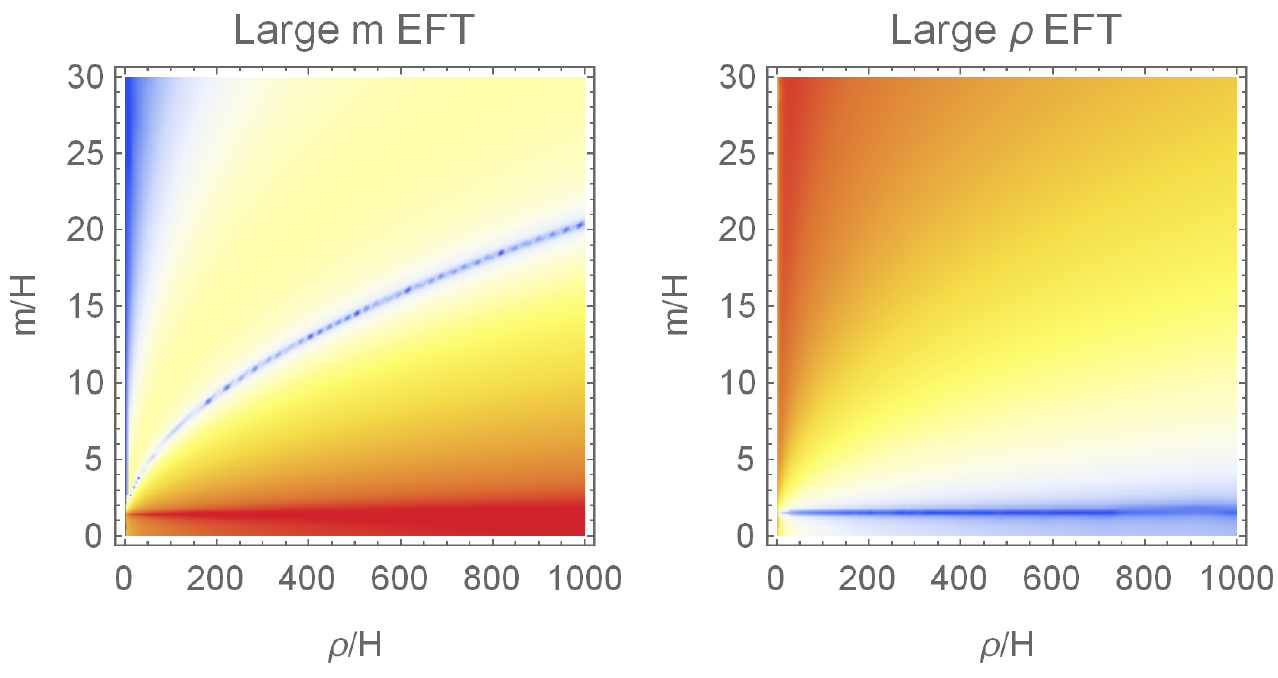}  
    \end{minipage} %
    \begin{minipage}{0.18\textwidth}
        \centering
        \includegraphics[width=0.6\linewidth]{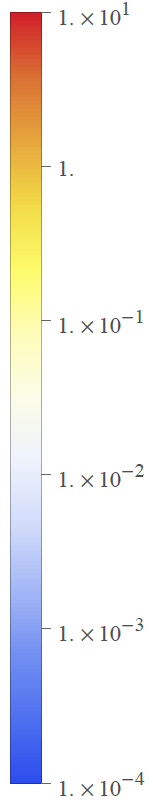}  
    \end{minipage}
  \caption{  A comparison of the relative error $( P_\zeta^{\rm EFT} - P_{\zeta}^{\rm full}) /P^{\rm full}_\zeta$ of the improved EFT, numerical EFT, large $\rho$ EFT and large $m$ EFT. We compare them with numerically solving the coupled differential equations which we regard as the exact value.} \label{Pzetadens}
\end{figure}

\begin{figure}[htbp] 
  \centering
  \includegraphics[width=0.8\textwidth]{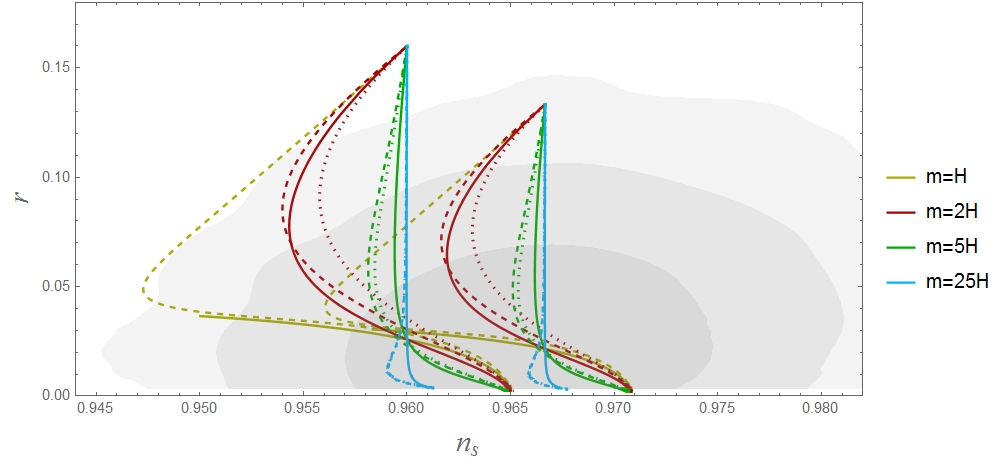}   
  \caption{ \label{nsr} Comparison of the analytical approximation of the improved EFT (solid line), numerical solution of the improved EFT (dotted), and the numerical result from solving the coupled linear differential equations (dashed) on the $n_s$-$r$ diagram. The yellow, red, green and blue correspond to the mass $m = H, 2H, 5H$ and $25 H$, respectively. The shaded region, from inner to outside, is the Planck 2015 $\sigma, 2\sigma$ and $3\sigma$ constraints on the $n_s$-$r$ diagram. There are two clusters of plots. The cluster on the left hand side corresponds to e-folding number $N=50$. The cluster on the right hand side corresponds to the e-folding number $N=60$.} 
\end{figure}

\begin{figure}[htbp] 
  \centering 
   \includegraphics[width=0.8\textwidth]{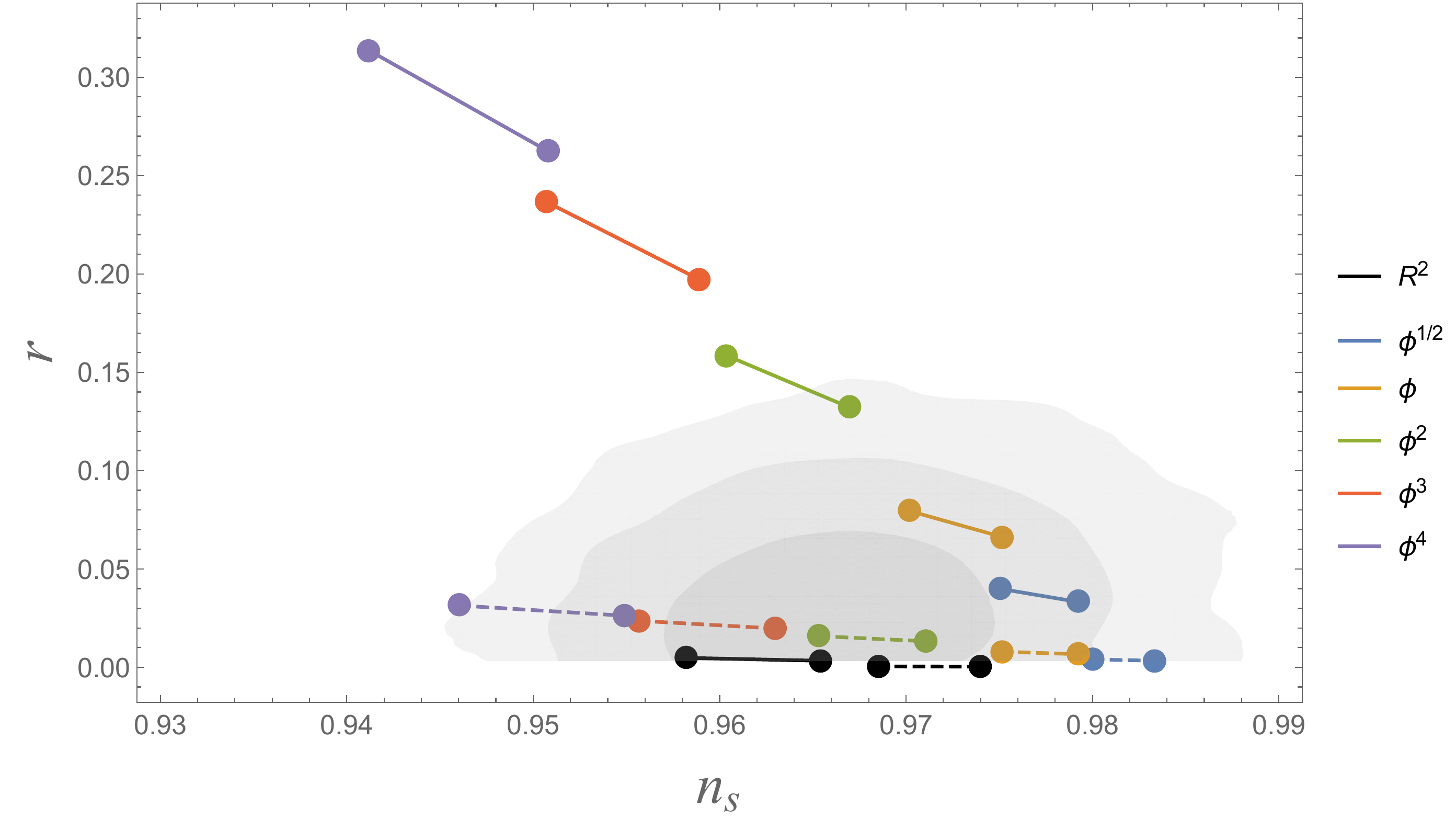}   
  \caption{\label{nsr1}  The shift of the prediction $n_s$-$r$ for different models. The solid lines correspond to the $n_s$-$r$ of the original single field model. The dashed lines correspond to the shifted $n_s$-$r$ values in the large $\rho$ limit. In this plot, we take $c_s=0.1$ (assuming mass $m<5H$) which is consistent with the experiment. Each line connects two dots. The dots on the left hand side corresponds to the e-folding number $N = 50$ and the dots on the right hand side corresponds to the e-folding number $N = 60$. The black, blue, yellow, green, red, purple lines correspond to the Starobinsky, $\phi^{1/2}$, $\phi$, $\phi^2$, $\phi^3$ and $\phi^4$ theories, respectively. As we can see from this plot, each model has a smaller $r$ value and a larger $n_s$ value after considering quasi-single field corrections.} 
\end{figure}

\begin{figure}[htbp] 
  \centering
  \includegraphics[width=0.7\textwidth]{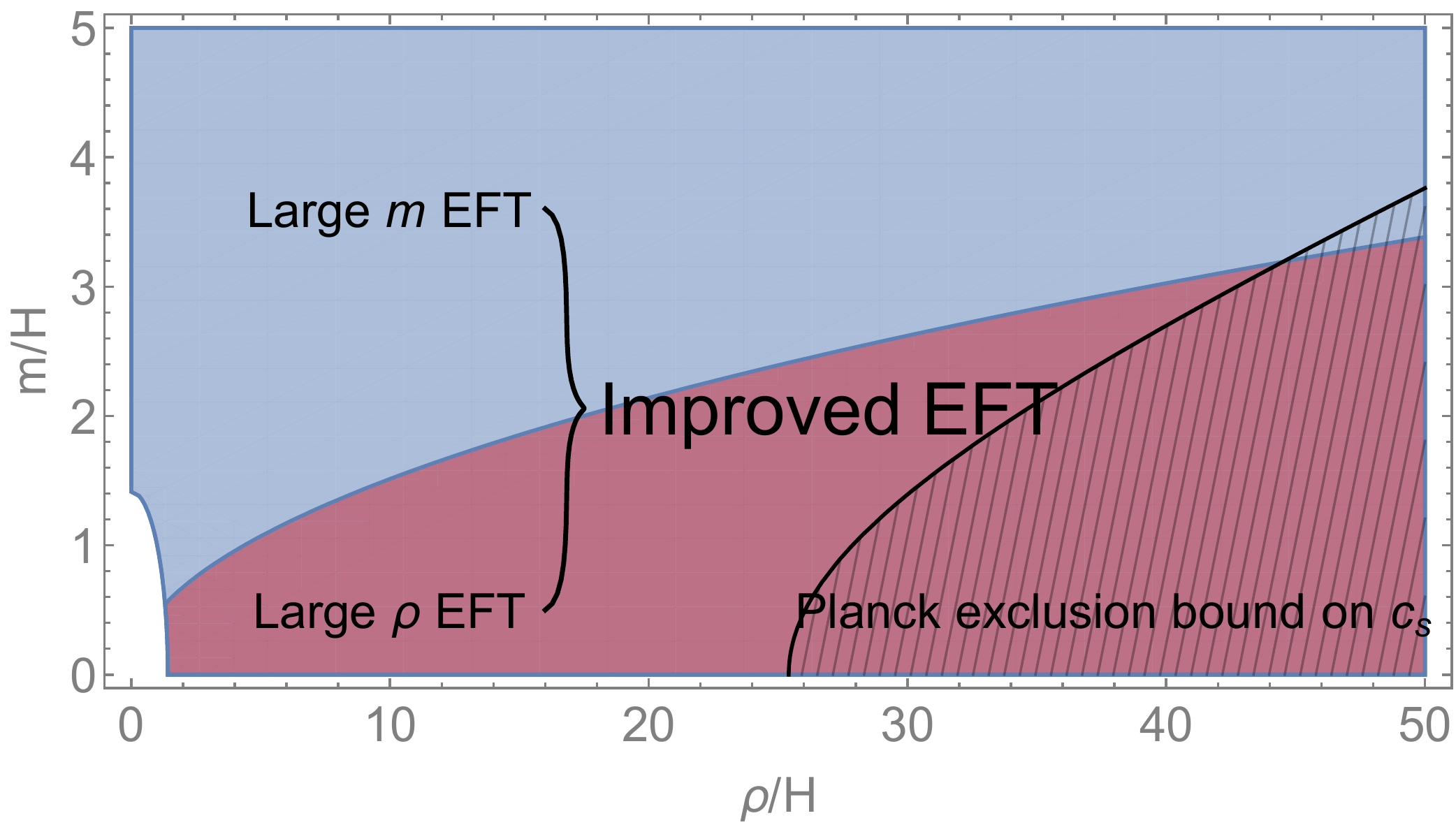}  
  \caption{\label{cs}  The parameter regime suitable for the large $m$ EFT ($\sqrt{\rho/H}<\mathcal C m/H$, blue) and large $\rho$ EFT ($\sqrt{\rho/H}> \mathcal C m/H$, red). The improved EFT method is valid for all the parameter regions satisfying $m^2+\rho^2>4H^2$. The white region with small $m$ and small $\rho$ ($m^2+\rho^2<4H^2$) is not covered by our improved EFT result. The Planck exclusion bound on the sound speed $c_s$ is also plotted in the figure as a shaded region. } 
\end{figure}

Next, we compare the large $m$ EFT, large $\rho$ EFT, improved EFT and numerical EFT with the exact value of numerically solving the coupled differential equations~\eqref{couple}. We plot the result on four density plots presented in Fig.~\ref{Pzetadens}. As we can see from the plots, large $m$ EFT is only suitable when $\sqrt{\rho/H}\ll m/H$ while large $\rho$ EFT only suitable when $\sqrt{\rho/H}\gg m/H$. The improved EFT can cover the region $\sqrt{\rho/H}\sim m/H$ with about 10\% error. The numerical EFT has a maximum error of 5\% and an average error of 0.1\% in the region that this method is applicable. The result indicates that in this region $m>\sqrt{2} H$ the EFT method is suitable and the real particle production contribution is negligible.

We continue to consider the shift on the $n_s$-$r$ diagram with the presence of the massive field. We adopt the definition of slow roll parameters as
\begin{align}
\epsilon = - \frac{\dot{H} }{H^2}, \quad \eta = \frac{\dot \epsilon}{\epsilon H}~.
\end{align}
The spectral index $n_s$ is defined through the scale dependence of the power spectrum
\begin{align}\nonumber
n_s -1 & \equiv \frac{d \ln P_{\zeta}}{d\ln k} = \frac{d\ln \left( \frac{H^2}{\epsilon} c_s^{-1} \right)}{d\ln k} = \frac{d\ln \left( \frac{H^2}{\epsilon} \right) + d\ln c_s^{-1} }{ d\ln k } \\ \nonumber
& = -2\epsilon -\eta + \frac{d\ln c_s^{-1}}{d\ln k} =  -2\epsilon -\eta + \frac{\partial \ln c_s^{-1}}{\partial \ln \rho} \frac{d\ln \rho}{d\ln \epsilon} \frac{d\ln \epsilon}{d N} \frac{dN}{d\ln k} +\frac{\partial \ln c_s^{-1}}{\partial \ln m } \frac{d\ln m}{d\ln H} \frac{d\ln H}{d N} \\
& = -2\epsilon - \eta  + \frac{1}{2} \frac{\partial \ln c_s^{-1}}{ \partial \ln \rho}  \eta + \frac{\partial \ln c_s^{-1}}{\partial \ln m} \epsilon~. 
\end{align}
In the expression of $n_s-1$, on the one hand, the errors in the power spectrum calculation (due to analytical approximation or numerical error) are amplified because of differentiation. On the other hand, the errors are multiplied by $\epsilon$ and $\eta/2$, which are at percent level or less. 

In the large $\rho$ scenario, $c_s^{-1}$ becomes independent of the mass of the $\sigma$ field, and the scaling of $c_s^{-1}$ is driven to $(\rho/H)^{1/2}$. As a result, the correction converges to
\begin{align}
n_s - 1 \xrightarrow{\rm large\,\, \rho} -2\epsilon - \frac{3}{4} \eta~.
\end{align}
Thus in the strong coupling limit, $n_s$ is attracted to a universal value for arbitrarily heavy massive fields (but with $m/H\lesssim \sqrt{\rho/H}$). We can write down the eventual spectral index for a given single field inflation model.  

\begin{align}
{\rm Starobinsky}: \qquad n_s = 1- \frac{9}{2 N^2} - \frac{2}{N} & \quad\Rightarrow\quad n_s = 1 - \frac{15}{4 N^2} - \frac{3}{2N} \\
\phi^n: \qquad n_s = 1-\frac{2n+4}{n} \epsilon_V \, & \quad\Rightarrow\quad n_s = 1 - \frac{2 n+3}{n} \epsilon_V, \quad \epsilon_V = \left( 1 + \frac{4 N}{n} \right)^{-1}
\end{align}
where $N$ is the e-folding number. For all these models, the correction to $n_s$ is positive. 

The power spectrum for the primordial gravitational waves is not influenced by the interactions between the inflaton and the massive fields
\begin{align}
P_{\gamma} = \frac{2}{\pi^2} \frac{H^2}{M_{\rm pl}^2}~.
\end{align}
The tensor to scalar ratio can be written as
\begin{align}
r = \frac{P_{\gamma}}{P_{\zeta}} = 16 \epsilon \times c_s ~.
\end{align}
As the coupling strength between the inflaton and the massive fields increases, the tensor to scalar ratio $r$ decreases indefinitely.

We plot the running trajectory of the $\phi^2$ inflation model prediction on the $n_s$-$r$ diagram in Fig.~\ref{nsr}. The numerical EFT result coincides with full theory solution very well if the mass is large enough (say, $5H$ or more). This again confirms the feasibility of ignoring heavy particle production and considering only the virtual processes, if viewed perturbatively. On the other hand, the analytical result for improved EFT is accurate when $\rho$ is big or $m$ is big (actually also in the small $m$ regime, i.e. $H\lesssim m\lesssim 2H$, where the improved EFT trajectory tracks the exact one considerably closely).

We select a few typical inflation models with a massive field and examine their behaviors in the large $\rho$ limit. The change of the $n_s$-$r$ diagram is plotted in Fig.~\ref{nsr1}. The simplest $m^2\phi^2$ model becomes favored by experiments again after considering quasi-single field corrections. This behavior is previously noticed by \cite{An:2017hlx}. Note that for the $m^2\phi^2$ model to be driven to the observationally favored regime in the $n_s$-$r$ diagram, $c_s$ should be considerably smaller than 1. Thus unless fine-tuned, one expects that the future non-Gaussianity experiments have the potential to test the consistency of such quasi-single field corrections.

The Planck constraint on the sound speed is $c_s>0.087$ \cite{Ade:2015ava}.  We derived a constraint on $\rho/H$ and $m/H$,
\begin{align}
(\rho/H)^2 - 131.118 (m/H)^2 < 642.026 ~.
\end{align}

The parameter region of the large $\rho$ EFT, large $m$ EFT and our improved EFT with the Planck exclusion region from the constraint on $c_s$ is plotted in Fig.~\ref{cs}. As we can see from this result, the large $\rho$ region is excluded from the Planck constraints on $c_s$ because when $\rho$ is large, the sound speed scales as $c_s\sim \sqrt{\rho/H}$. Also although there is no explicit constraint from the $n_s$-$r$ diagram, as we can see from Fig.~\ref{nsr}, in general, bigger $\rho/H$ is favored by data for $m^2\phi^2$ inflation. This further motivates us to consider the intermediate region of $\sqrt{\rho/H}\sim m/H$, which is a more favored parameter region from both the constraints from the sound speed and the constraint from the $n_s$-$r$ diagram. 

\section{Conclusion}
\label{sec:conclusion}
We study the EFT for quasi-single field inflation, taking into account of all the virtual particle contributions. In the analytical approximation, an effective sound speed is introduced. The sound speed is a function of the mass of the massive particle $m$, the coupling $\rho$, and the time scale $k\tau$. Previously the power spectrum is only obtained in the large $m$ limit ($m/H\gg\sqrt{\rho/H}$) and the large $\rho$ limit ($\sqrt{\rho/H}\gg m/H$). We used a horizon freezing approximation method to estimate the sound speed and the primordial power spectrum. Because of the nice property of the inflaton that it will freeze at the time scale $k\tau\sim c_s^{-1}$. The effective sound speed can be solved. Our result can cover all the parameter region where an EFT can be applied, including the region $\sqrt{\rho/H}\sim m/H$ with an error of 10\% compared with numerically solving the coupled differential equations. This corresponds to an error of 0.1\% in the $n_s$-$r$ diagram.

We further check how big the real particle production process contributions to the power spectrum by comparing numerically solving the single equation of motion of the $\pi$ field and numerically solving the coupled differential equations. We find that in the dominate parts of the parameter space (for large $m$, $\rho$ parameters), the real particle production accounts for 0.1\% of the primordial power spectrum thus can be neglected in comparing power spectra observables with observation.

\section*{Acknowledgments}
We thank Henry Tye, Sam Wong and Gianluca Zoccarato for discussion. This work was supported by ECS Grant 26300316 and GRF Grant 16301917 from the Research Grants Council of Hong Kong. XT is supported by the 2017 Summer Research Program of the Hong Kong University of Science and Technology and the Qian San-Qiang Class in the University of Science and Technology of China. SZ is supported by the the Hong Kong PhD Fellowship Scheme (HKPFS) issued by the Research Grants Council (RGC) of Hong Kong.

\appendix
\section{A Brief Review about large $\rho$ EFT and large $m$ EFT}\label{largeEFT}

We review the previously known limits of the quasi-single field EFT.

$\bullet$ Large $\rho$ limit \cite{Baumann:2011su,Cremonini:2010ua,Gwyn:2012mw,An:2017hlx}: The effective action becomes 
\begin{align}
S_{\rm eff} = \frac{1}{2} \int \frac{d^3 k}{(2\pi)^3} d\tau \frac{1}{H^2\tau^2} \bigg[ \frac{\rho^2}{H^2\tau^2k^2 } ( \partial_\tau \pi )^2 -k^2 \pi^2 \bigg]~.
\end{align}
The conjugate momentum of the $\pi$ field is
\begin{align}
\theta = \frac{\delta S}{\delta \pi'}  = \frac{\rho^2}{H^4\tau^4 k^2} \pi'~.
\end{align}
The $\pi$ field is quantized as
\begin{align}
\pi_{\mathbf k} (\tau) = u_{k} (\tau) a_{\mathbf k} + u^*_{k} (\tau) a_{-\mathbf k}^{\dagger}~.
\end{align}
The $\pi$ field and its conjugation $\theta$ should satisfy the commutation relation
\begin{align}
[\pi_{\mathbf k} (\tau), \theta_{\mathbf k'} (\tau) ] = i (2\pi)^3 \delta^3 (\mathbf k+\mathbf k')~,
\end{align}
which implies that the mode function $u_k(\tau)$ and its conjugate $u_{k}^*(\tau)$ should satisfy the following normalization
\begin{align}
(u_k u'^*_k - u'_k u^*_k) \frac{\rho^2}{H^4\tau^4 k^2} = i~.
\end{align}
The equation of motion for the $\pi$ field is 
\begin{align}
\frac{\rho^2}{H^2} \frac{d}{d ( k\tau )} \left( \frac{1}{k^4\tau^4} \frac{du_k}{d(k\tau)} \right) + \frac{u_k}{k^2\tau^2} = 0~.
\end{align}
Using the normalization condition to fix the coefficients, the solution of the $\pi$ field becomes
\begin{align}
u_k (\tau) = \left( \frac{2\pi^2\rho}{H} \right)^{1/4} \frac{H}{\sqrt{2 k^3}} \left( \frac{k^2\tau^2 H}{2\rho} \right)^{5/4} H_{5/4}^{(1)} \left( \frac{k^2\tau^2 H}{2\rho} \right)~.
\end{align}
The power spectrum is 
\begin{align}
P_{\zeta} = \frac{H^2}{\dot \phi^2} |u_k(\tau)|^2_{\tau\rightarrow 0} = \frac{H^4}{\dot \phi^2} \left( \frac{1}{2 k^3} \right) \bigg[ \frac{16\pi}{\Gamma^2(-1/4)} \left( \frac{\rho^2}{H} \right)^{1/2} \bigg]~.
\end{align}
It leads to the correction to the original power spectrum as
\begin{align}
P_{\zeta} =  P_{\zeta}^{(0)} (k) c_s^{-1}, \quad c_s^{-1} = \mathcal C \bigg( \frac{\rho}{H} \bigg)^{1/2}~,
\end{align}
where $\mathcal C = 16\pi/\Gamma^2(-1/4)\simeq 2.09$ is a constant.

$\bullet$ Large $m$ limit \cite{Tolley:2009fg,Achucarro:2010jv,Achucarro:2012sm}: The effective action in the large $m$ limit is
\begin{align}
S_{\rm eff} = \frac{1}{2} \int \frac{d^3 k}{(2\pi)^3} d\tau \frac{1}{H^2\tau^2} \bigg[ \left(1+\frac{\rho^2}{m^2- 2 H^2} \right) (\partial_\tau \pi)^2 - k^2 \pi^2 \bigg]~.
\end{align}
The conjugate momentum is
\begin{align}
\theta = \frac{\delta S }{\delta \pi'} = \frac{1}{H^2\tau^2} \left(1+\frac{\rho^2}{m^2- 2 H^2} \right) \pi'~.
\end{align} 
The normalization is
\begin{align}
(u_k u'^*_k - u'_k u^*_k) \left( 1+\frac{\rho^2}{m^2- 2H^2} \right) = i~.
\end{align}
After taking this normalization into account, the solution to the mode function $\pi$ is
\begin{align}
u_k (\tau) = \frac{H}{\sqrt{2 c_s k^3}} (1+ i c_s k \tau) e^{-i c_s k \tau}, \quad c_s = 1/\sqrt{1+\frac{\rho^2/H^2}{m^2/H^2- 2}}~.
\end{align}
As a side remark that a similar result $c_s = 1/\sqrt{1+\frac{\rho^2/H^2}{m^2/H^2-9/4}}$ can be obtained by explicit summing over the Feynman diagrams using the Schwinger-Keldysh formalism.

\end{document}